# *Instrumental genesis through interdisciplinary collaboration – reflections on the emergence of a visualisation framework for video annotation data*

XML

This paper presents, discusses and reflects on the development of a visualization framework for the analysis of the temporal dynamics of audiovisual expressivity. The main focus lies on the instrumental genesis process (Rabardel 1995; Longchamp 2012) – a concept trying to express and analyze the co-evolution of instruments and the practices they make possible – underlying this development. It is described through the collaboration and communication processes between computer science scholars and humanities scholars in finding new ways of visualizing complex datasets for exploration and presentation in the realm of film-studies research. It draws on the outcome and concrete usage of the visualizations in publications and presentations of a research group, the AdA-project, that investigates the audiovisual rhetorics of affect in audiovisual media on the financial crisis (2007–). These film analyses are based on theoretical assumptions on the process of film-viewing, the relation of the viewer's perception and the temporally unfolding audiovisual images, and a methodical approach that draws on 'steps' in the research process such as segmentation, description and qualification, called eMAEX (Kappelhoff et al. 2011–2016) to reconstruct these experiential figurations (Bakels et al. 2020a, 2020b).

The main focus of this paper is the process of iterative development of visualizations as interactive interfaces generated with the open-source software Advene, that were an integral part of the research process. In this regard, the timeline visualization is not only of interest for visual argumentation in (digital) humanities publications, but also for the creation of annotations as well as the exploration of this data. In the first part of the paper we describe this interdisciplinary collaboration as instrumental genesis on a general level – as an evolving and iterative process. In the second part we focus on the specific challenge of designing a visualization framework for the temporal dynamics of audiovisual aesthetics. Lastly we zoom out by reflecting on experiences and insights that might be of interest for the wider digital humanities community.

## 1. Growing together: evolving tools for evolving concepts

In the AdA-project the research process encompassed different steps such as creating annotation data (either computational or manually), exploring the data related to moving images, finding patterns and reconstructing figurations as well as communicating them to others. For the explorative comparative study of an extensive corpus, large data sets of detailed film-analytical, systematic, machine-readable annotations have been created. In this process, the Advene software has been involved at the various stages, from the exploration of data modelling and visualizations and the actual creation of annotation datasets, to the exploration of annotations in order to support analyses. Advene has been designed as a tool for audiovisual active reading (Aubert / Prié 2005; Aubert et al. 2012), allowing users to create and share annotation-based interpretations of audiovisual media. In order to become a part of the creative scholarly processes involved in the development of media analyses, it is built upon flexible data structures and visualizations. In the course of a scholarly activity, as underlined by (Prié 2011), information structures are constantly evolving to various extents. Moreover, these evolving data structures are mediated by visualizations, that also have to evolve in order to match the needs of both structure evolution and thought process. The modalities and temporalities of these evolutions depend on the possibilities offered by the software used to mediate the information, and on the skills of the users: some evolutions of data or visualizations can consist in simple or more advanced configurations of existing tools or visualizations, that can be carried out autonomously up to a certain point. For more advanced evolutions of data structure and visualizations, more methodological and technological know-how may be involved and benefit from a collaboration with computer science scholars, who can provide insights and advice in modelling and configuration as well as possibly develop new tools when needed. The cooperation between computer science scholars and humanities scholars can neither be conceptualized as the ordering of a fixed product that fulfills pre-existing needs of humanities scholars, nor the deployment of state-of-the-art visualization modes by computer science scholars that are then just applied. In our experience, as can be also seen in (El Kathib 2019), interdisciplinary research in the realm of digital humanities rather requires joint processes of sketching out needs, possibilities and restrictions and translating affordances into the languages of both domains. Such translation requires inter-comprehension, which can be achieved through mutual appropriation of the domain-specific concepts, or at least have some members "able to speak across various disciplines" (Siemens et al. 2010). The 'product' of such a development process is thus more aptly described as a framework that enables scholars to think dynamically within a certain data-visualization paradigm.

## 2. Mapping time and movement: the development of a visualization framework for media studies

We ground these reflections on the evolution of timeline visualizations, that was used throughout the project, describing why and how it had to evolve in order to support the analyses of the humanities scholars, as well as the collaboration process that led to these evolutions. We will also underline some of the design principles that offer a level of empowerment to humanities scholars in adapting knowledge instruments.

The challenge of creating visualizations of the temporal dynamics of audiovisual media has been approached in the realm of film studies from different angles and with diverging research interests: from the hand-drawn film scores by Russian formalists like Sergei Eisenstein (Eisenstein 2010 [1941]) that focus on the synchronicity of various filmic modes to its digital reimagination (Heftberger 2018) and digital frameworks that focus on specific aesthetic dimensions like color (Flückiger / Halter 2020) or montage rhythm (Tsivian 2009). The main challenges in our case were the quantity of metadata, its semantic structure (Agt-Rickauer et al. 2018), the need for a systematized mode of visualization for a larger corpora through multiple annotators, while at the same time acknowledging the variance within the corpus. Even within a theoretical framework there is no one-fits-all-solution for visualizing the temporal dynamics of different films. Whereas in one sequence the cutting rhythm and character movements might be crucial, other sequences dominantly use music and changes in the color scheme to convey moods and meanings. Therefore the timeline visualizations are designed as sandboxes, that allow scholars without programming skills to rearrange, reshape and adjust the visual output on the fly within a restricted set of possibilities.

Adding configurability to a digital instrument actually increases its complexity from both the user's and the designer's point of view. The conception of a configurable instrument requires the creation of both the instrument itself, and also of its configuration means, which both define the possibilities it offers. In our case, the first version of the timeline (see Figure 1) was the one provided as a standard interface by Advene. It provides a generic way to interact with the annotations, both for creation and exploration. It offers different levels of configuration, which have been exploited during the project: the ability to define the presented types of annotations, its behavior when playing the associated video, etc. Some of the parameters like advanced color specification based on annotation content are accessible through a specific language syntax, that had to be appropriated by the humanities scholars in order to produce some variations in visualization, exhibiting characteristics of the data. However, the interaction facilities that the timeline offers can reduce its capacity to handle a larger number of annotations, and in the course of the project, the interdisciplinary discussions, exchanging requirements and constraints from both humanities and digital domains, led to the conclusion that another visualization component was needed. A new version of the timeline has thus been designed (see Figure 2 for a sketch used during the discussion phase, and Figure 3 for the new version), focusing more on data exploration than data creation and manipulation. One of the design features of the new version is the syntax used for describing the configuration of the timeline: it is a text-based (actually encoded as a human-readable URL) syntax, akin to a Domain-Specific Language (DSL) fitted for the visualization of the specific data structure of the AdA-project. This syntax presents a bit more complexity to use compared to a full-fledged graphical user interface, but it also offers more flexibility, allowing scholars to do independent variations and explorations with different settings. Moreover, its textual characteristic makes it more apt at being saved for further references or sharing, hence improving the reproducibility of the explorations. Additionally, not having to develop a full-fledged configuration interface, this approach allows to dedicate development time on core functionalities, in the constant tradeoff of finding a balance between the time gain offered by using a new feature and the time needed to develop it.

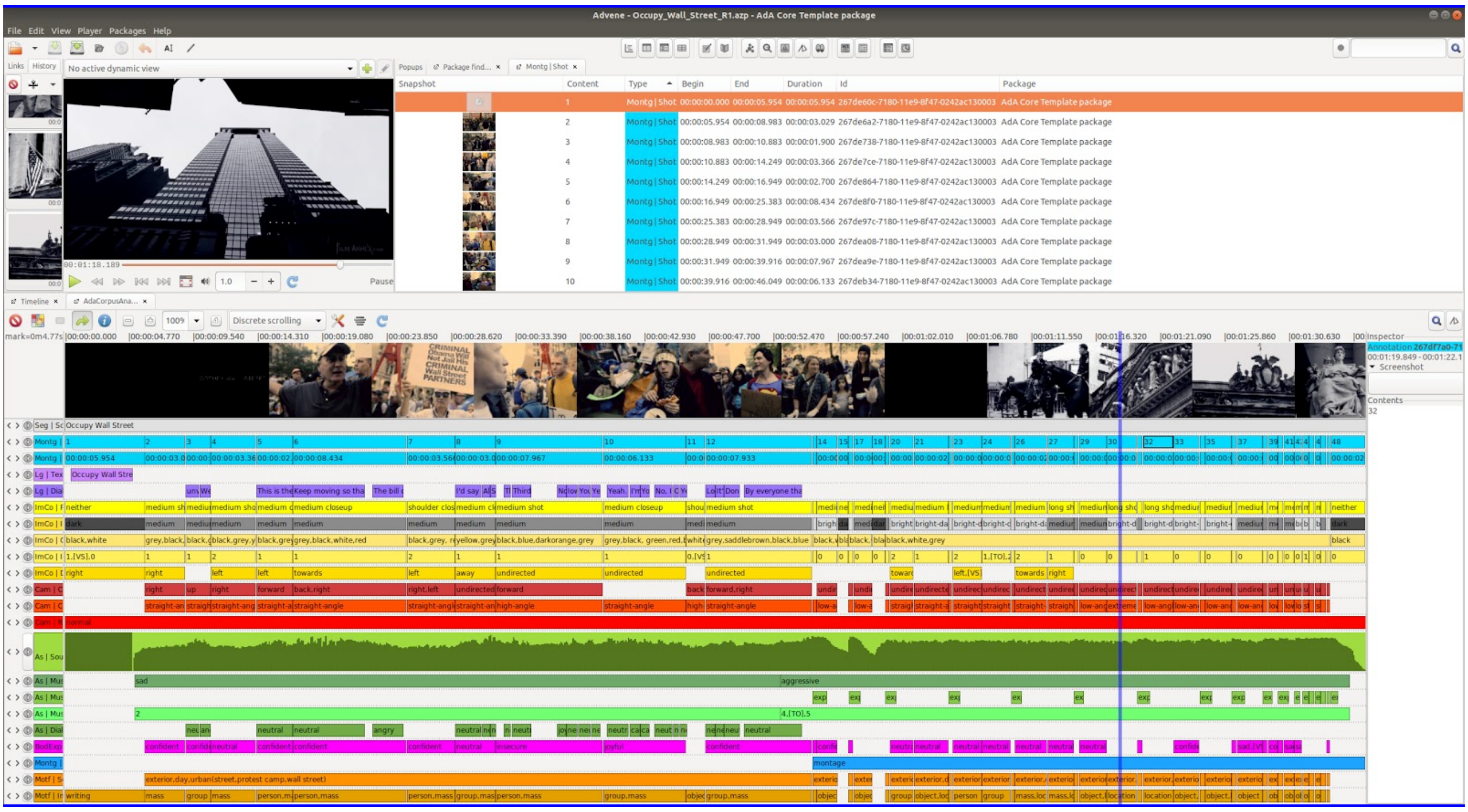

Fig. 1: Pre-existing timeline editing interface in Advene

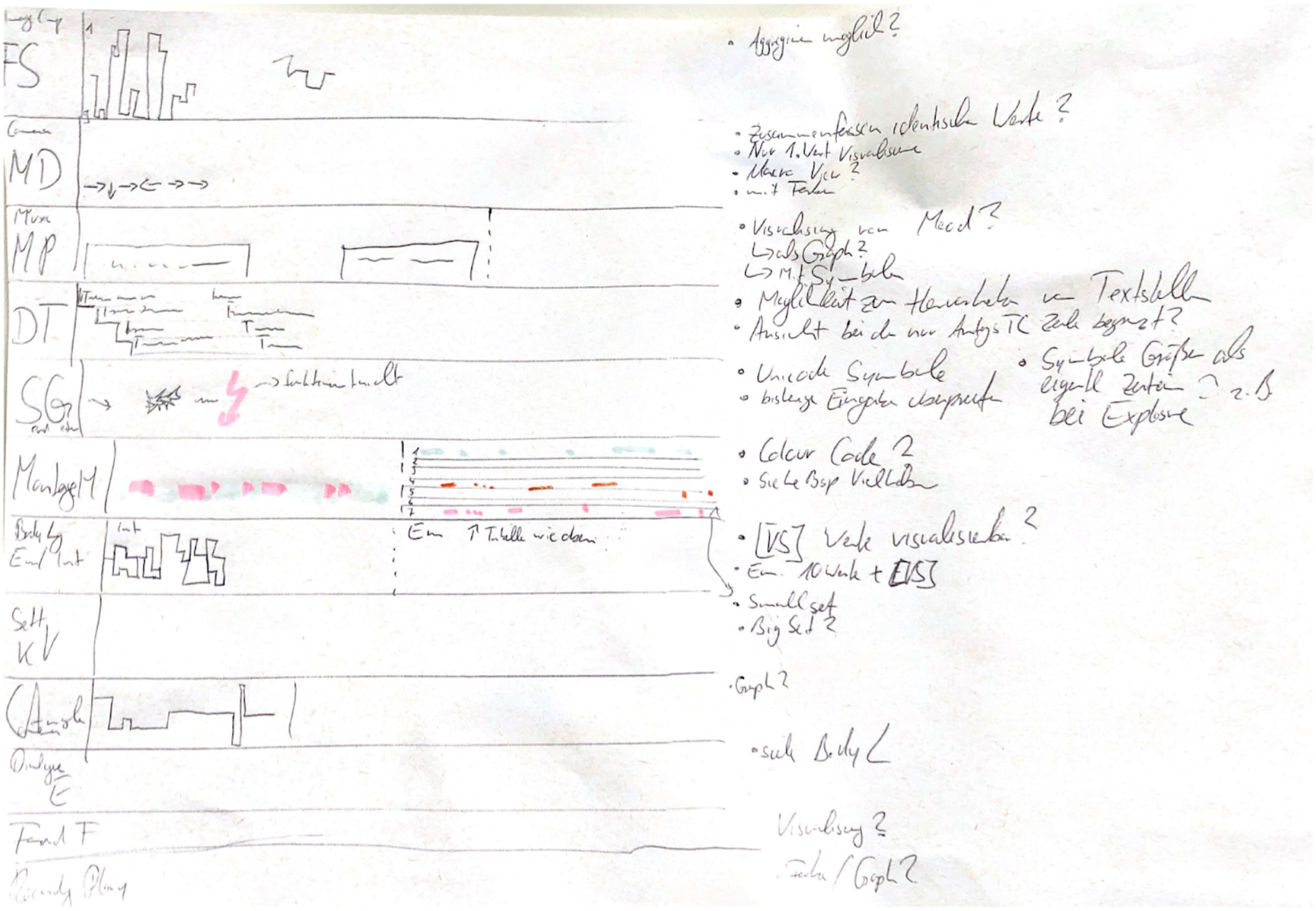

Fig. 2: Sketch used during the new timeline creation process

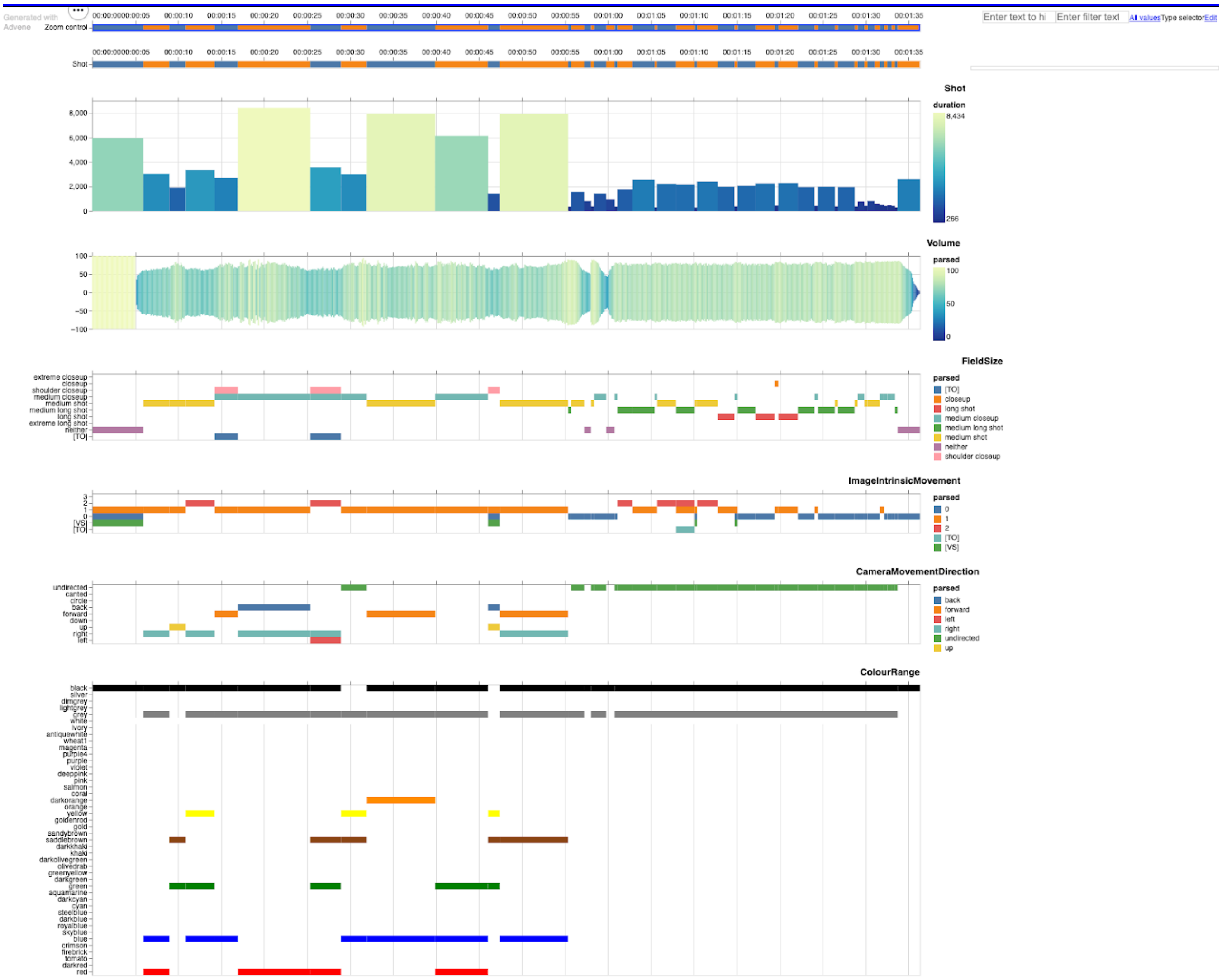

Fig. 3: New web-based version of the timeline for exploration and presentation

## 3. Reflections on an iterative & interdisciplinary process

These concrete procedures and considerations open up a discourse about the conceptualization of “software solutions” in the realm of digital humanities in a more general sense. Between the notions of ‘tool’ and ‘instrument’, the interdisciplinary process of the development appears as a manyfold and difficult process of crafting, appropriation, enablement and exploration. By reflecting on the instrumental genesis of a visualization framework, we aim in this paper at highlighting the process rather than the outcome. Especially in the realm of digital humanities these processes of inter- and transdisciplinary communication and exchange pose a major challenge for many research projects – especially when they work with challenging and complex data types. By mapping out the problems and possibilities that arise from the friction and multitude of perspectives not as an abstract state but as organic parts of a concrete development process, we try to bridge the gap between experience report and theoretical self-reflection

## Appendix A

Bibliography

1. Aubert, Olivier / Prié, Yannick (2005): "Advene: Active Reading through Hypervideo", in: *Proceedings of the sixteenth ACM conference on Hypertext and hypermedia* , sep. 2005, 235–244 DOI: 10.1145/1083356.1083405 .
2. Aubert, Olivier / Prié, Yannick / Schmitt, Daniel (2012): “Advene as a Tailorable Hypervideo Authoring Tool: a Case Study”, in: *Proceedings of the 2012 ACM symposium on Document engineering* , sep. 2012, 79–82 DOI: 10.1145/2361354.2361370 .
3. Agt-Rickauer, Henning / Aubert, Olivier / Hentschel, Christian / Sack, Harald (2018): “Authoring and Publishing Linked Open Film-Analytical Data”, in: EKAW 2018 Posters and Demonstrations Session , Nancy, nov. 2018.
4. Bakels, Jan-Hendrik / Grotkopp, Matthias / Scherer, Thomas / Stratil, Jasper (2020a): "Digitale Empirie? Computergestützte Filmanalyse Im Spannungsfeld Von Datenmodellen Und Gestalttheorie.", in: montage/AV 29, 1, 99–118.
5. Bakels, Jan-Hendrik / Grotkopp, Matthias / Scherer, Thomas / Stratil, Jasper (2020b): "Matching Computational Analysis and Human Experience. Performative Arts and the Digital Humanities.", in: Digital Humanities Quarterly 14, 4.

http://www.digitalhumanities.org/dhq/vol/14/4/000496/000496.html [26.05.2021].

6. Eisenstein, Sergei M. (2010 [1941]): “Vertical Montage”, in: Michael Glenny, Richard Taylor (eds.) Sergei Eisenstein. Selected Works. Volume 11. Towards a Theory of Montage. New York, I.B.Tauris, 327–399.

7. El Khatib, Randa / Wrisley, David Joseph / Elbassuoni, Shady / Jaber, Mohamad / El Zini, Julia (2019): “Prototyping Across the Disciplines”, in: Digital Studies/Le champ numérique , 8, 1, p.10. DOI: 10.16995/dscn.282 .

8. Flückiger, Barbara / Halter, Gaudenz (2020): "Methods and Advanced Tools for the Analysis of Film Colors in Digital Humanities.", in: Digital Humanities Quarterly 14, 4. http://www.digitalhumanities.org/dhq/vol/14/4/000500/000500.html [26.05.2021].

9. Heftberger, Adelheid (2018): Digital Humanities and Film Studies. Visualising Dziga Vertovs’s Work . Cham, Springer VS. DOI: 10.1080/00401706.2019.1679533 .

10. Kappelhoff, Hermann / Bakels, Jan-Hendrik / Berger, Hanno / Brückner, Regina / Böhme, Dorothea / Chung, Hye-Jeung / Dang, Sarah-Mai / Gaertner, David / Greifenstein, Sarah / Gronmaier, Danny / Grotkopp, Matthias / Haupts, Tobias / Illger, Daniel / Lehmann, Hauke / Lück, Michael / Pogodda, Cilli / Roleff, Naomi / Rook, Stefan / Rositzka, Eileen / Scherer, Thomas / Schlochtermeier, Lorna / Schmitt, Christina / Steininger, Anna / Tag, Susanne (2011–2016): "Empirische Medienästhetik. Datenmatrix Kriegsfilm –eMAEX." https://www.empirische-medienaesthetik.fu-berlin.de/emaex-system/index.html [26.05.2021].

11. Lonchamp, Jacques (2012): “An instrumental perspective on CSCL systems”, in: International Journal of Computer-Supported Collaborative Learning , 7, 211–237.

12. Prié, Yannick (2011): Vers une phénoménologie des inscriptions numériques. Dynamique de l'activité et des structures informationnelles dans les systèmes d'interprétation. Université Claude Bernard - Lyon I, 2011. ⟨ tel-00655574⟩ .

13. Rabardel, Pierre (1995): Les hommes et les technologies, une approche cognitive des instruments contemporains , Armand Colin, Paris.

14. Siemens, Lynne / Cunningham, Richard / Duff, Wendy / Warwick, Claire (2010): “More Minds are Brought to Bear on a Problem”, in: Methods of Interaction and Collaboration within Digital Humanities Research Teams. Digital Studies/le Champ Numérique , 2, 2. DOI: 10.16995/dscn.80 .

15. Tsivian, Yuri (2009): “Cinemetrics, part of the humanities’ cyberinfrastructure”, in M. Ross, M. Grauer and B. Freisleben (eds.), Digital tools in media studies: Analysis and research , transcript, Bielefeld, 93–100

*Olivier Aubert (contact@olivieraubert.net), University of Nantes, France and Thomas Scherer (scherer.thomas@fu-berlin.de), Freie Universität Berlin and Jasper Stratil (jasper.stratil@fu-berlin.de), Freie Universität Berlin*